\documentclass[]{spie}  

\usepackage[ruled,vlined]{algorithm2e}
\usepackage{array}
\usepackage{multirow}
\usepackage{tabularx}
\usepackage{amsfonts}
\usepackage{algcompatible}
\usepackage{amsmath}
\usepackage{mathrsfs}
\usepackage{graphicx}
\usepackage{indentfirst}
\DeclareUnicodeCharacter{2061}{}

\usepackage[colorlinks=true, allcolors=blue]{hyperref}

\title{MB-DECTNet: A Model-Based Unrolled Network for Accurate 3D DECT Reconstruction}

\author[a]{Tao Ge}
\author[a]{Maria Medrano}
\author[a]{Rui Liao}

\author[b]{David G. Politte}
\author[c]{Jeffrey F. Williamson}
\author[d]{Bruce R. Whiting}
\author[a]{Joseph A. O’Sullivan}

\affil[a]{Washington University in St. Louis, Department of Electrical \& Systems Engineering}
\affil[b]{Washington University in St. Louis, Mallinckrodt Institute of Radiology}
\affil[c]{Washington University in St. Louis, Radiation Oncology, St. Louis, MO 63110}
\affil[d]{University of Pittsburgh, Department of Radiology, Pittsburgh, PA 15213}

\authorinfo{Further author information: \\Tao Ge: getao@wustl.edu \\  Joseph A. O’Sullivan: jao@wustl.edu}

\begin{document} 
\maketitle

\begin{abstract}

Numerous dual-energy CT (DECT) techniques have been developed in the past few decades. Dual-energy CT (DECT) statistical iterative reconstruction (SIR) has demonstrated its potential for reducing noise and increasing accuracy. Our lab proposed a joint statistical DECT algorithm for stopping power estimation and showed that it outperforms competing image-based material-decomposition methods. However, due to its slow convergence and the high computational cost of projections, the elapsed time of 3D DECT SIR is often not clinically acceptable. Therefore, to improve its  convergence, we have embedded DECT SIR into a deep learning model-based unrolled network for 3D DECT reconstruction (MB-DECTNet) that can be trained in an end-to-end fashion. This deep learning-based method is trained to learn the shortcuts between the initial conditions and the stationary points of iterative algorithms while preserving the unbiased estimation property of model-based algorithms. MB-DECTNet is formed by stacking multiple update blocks, each of which consists of a data consistency layer (DC) and a spatial mixer layer, where the spatial mixer layer is the shrunken U-Net, and the DC layer is a one-step update of an arbitrary traditional iterative method. Although the proposed network can be combined with numerous iterative DECT algorithms, we demonstrate its performance with the dual-energy alternating minimization (DEAM). The qualitative result shows that MB-DECTNet with DEAM significantly reduces noise while increasing the resolution of the test image. The quantitative result shows that MB-DECTNet has the potential to estimate attenuation coefficients accurately as traditional statistical algorithms but with a much lower computational cost. 
\end{abstract}


\section{INTRODUCTION}

Dual-energy computed tomography (DECT) has been widely investigated to generate informative and accurate images \cite{McCollough2015, Hyun2017}. Dual-energy CT (DECT) statistical iterative reconstructions (SIR) have demonstrated their potential to reduce noise and increase accuracy \cite{Zhang2014, Jin2015, Yong2014, Zhang_2017}. For instance, our lab proposed a joint statistical DECT algorithm, dual-energy alternating minimization (DEAM), which outperforms competing methods and achieves sub-percentage uncertainty in estimating proton stopping power \cite{O07,Zhang2018,Zhang2019,Maria2022}.

However, 3D DECT SIRs are usually time-consuming, even with multiple acceleration techniques, including multi-GPU acceleration \cite{Ayan2017} and ordered-subsets method \cite{Hudson1994}. Compared to single-energy CT (SECT), DECT algorithms reconstruct two measured sinograms scanned at two different spectra, which at least doubles the system operations per iteration. Moreover, since approximations are usually used in DECT SIRs to evaluate the gradient of the polychromatic forward models, DECT SIRs always converge much slower than single-energy algorithms. The low convergence rate and high computational cost of projections make it difficult to get an accurate DECT result within a clinically acceptable time.

To reduce the elapsed time while retaining the estimation accuracy, we incorporate DECT SIR into a deep learning model-based unrolled network for DECT reconstruction (MB-DECTNet). This work is based on deep unfolding \cite{Liu2021}, which simulates iterative algorithms using a series of convolutional neural networks (CNN) \cite{Hauptmann2018, Aggarwal2019}. We substitute the data consistency layer in deep unfolding with DECT gradients. Then, in MBDECTNet, each individual block can be considered an iteration of JSIDECT, while the step size and the image prior are determined by the CNN.

\section{Methods}
\subsection{Problem Description}
In contrast to single-energy CT, DECT estimates various types of images from two sinograms measured at different spectra. According to Beer’s law, the transmission data can be modeled as a Poisson random vector with the entry

\begin{equation}
d_j(y)\sim \texttt{Poisson}\Big\{\int I_{0,j}(y,E) \exp\Big(-\int_X h(x,y)\mu(x,E)dx\Big)dE\Big\},
\end{equation}
where $x\in X$ denotes spatial location or voxel index in the image domain, $y\in Y$ denotes the index (fan angle, gantry angle, and bed position) of the measurement, $j={L,H}$ denotes the spectrum, E denotes the energy, $I_0$ is the photon counts in the absence of an object, which contains information about the bowtie filter and spectrum, etc., $h(x,y)$ denotes the system operator, and $\mu(x,E)\in \mathbb{R}^+$ denotes the linear attenuation coefficient (LAC) of the object at location $x$ and energy $E$.
Traditional SIR algorithms solve the DECT material-decomposition problem by iteratively minimizing the objective function \cite{Zhang2014,O07,Jin2015,Yong2014}
 \begin{equation}
\arg\min_{\boldsymbol{c}}⁡\sum_j\sum_y\mathscr{D}\{d_j(y),g_j(y:\boldsymbol{c})\}+\mathscr{R}(\boldsymbol{c}),
\end{equation}
where $\mathscr{D}:(Y,Y)\to \mathbb{R}$ and $\mathscr R:X\to \mathbb{R}$ denote the data fidelity and penalty term, respectively; $\boldsymbol{c}=\{c_i\}$ denotes image of basis weights, and $i$ indexes over basis materials, $g_j (y:c)$ is the sinogram predicted by the discretized forward model operating on the current basis-material-decomposition, $c$
\begin{equation}
g_j (y:\boldsymbol{c})=\sum_E I_{0,j}(y,E) \exp\Big(-\sum_x h(x,y)\sum_i c_i(x)\mu_i(E)\Big),
\end{equation}
where $\mu_i (E)$ is the LAC of the $i$-th basis material at energy $E$.

Prior work from our lab shows that DECT SIR algorithms can reconstruct images with subpercentage accuracy from uncorrected transmission data  \cite{Zhang2018,Zhang2019,Maria2022}. However, the elapsed time of statistical algorithms for 3D DECT is usually time-consuming because of the high computational cost of the system operator as well as the low convergence rate. To address this issue, we introduce the unrolled network for 3D DECT reconstruction in the next subsection.

\subsection{Model-Based Unrolled Network}

\begin{figure}
  \centering
  \includegraphics[width=0.98\textwidth]{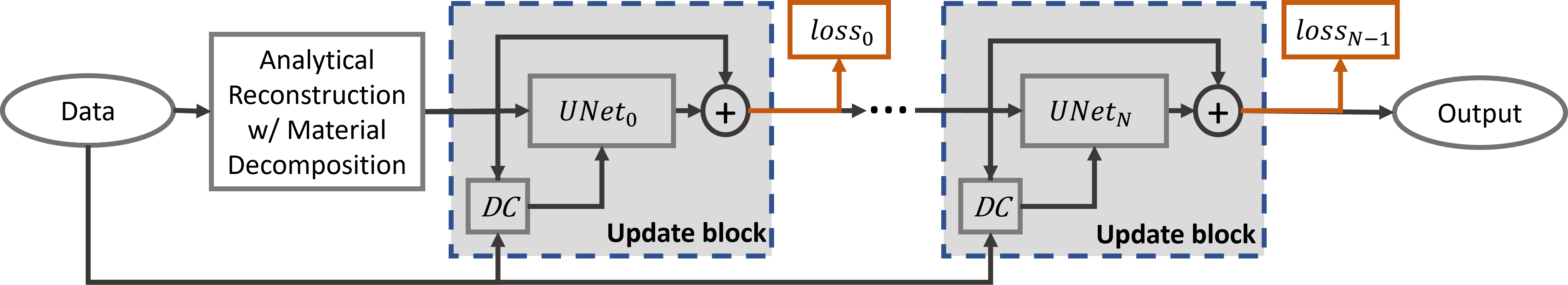}
  \caption{Flowchart of the proposed pipeline. Each blue dashed box denotes an update block, which mimics an iteration in DECT SIR. DC is the data consistency layer, and the orange boxes evaluate the loss functions for training.}
  \label{fig:pipeline}
\end{figure}

The basic idea of MB-DECTNet is using a set of stacked neural networks to simulate the iterative update process of the traditional SIR algorithms. Figure \ref{fig:pipeline} shows the flowchart of the proposed pipeline. The data is firstly reconstructed by an analytical algorithm combined with material decomposition to generate the initial condition for the network. The unrolled network consists of several update blocks, and each update block mimics a single iteration of the traditional statistical algorithm.

\begin{figure}
  \centering
  \includegraphics[width=0.98\textwidth]{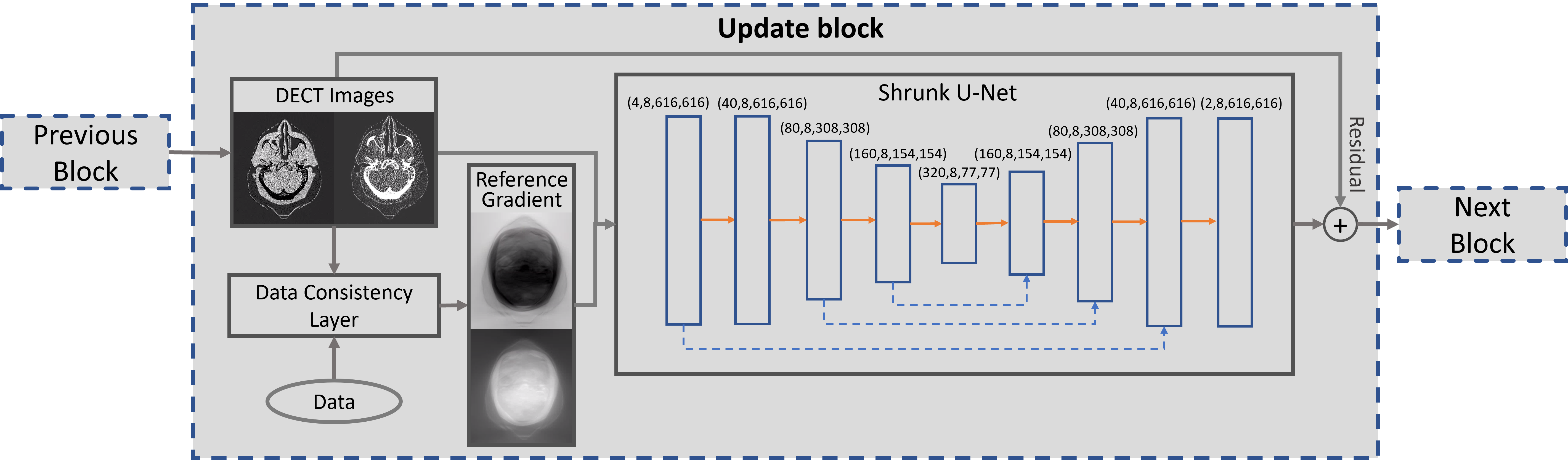}
  \caption{Structure of a single update block.}
  \label{fig:detailed}
\end{figure}

The structure of the update block is shown in Figure \ref{fig:detailed}. The data consistency (DC) layer takes the image from the previous block and the transmission data as the inputs, and outputs the reference gradient, defined as the difference between and the minimizer of the surrogate data fidelity term and the previous iterate $c^{k-1}$:

\begin{equation}
    \mathcal{DC}(\boldsymbol{c}_{k-1},d)=\Big\{\arg\min_{\boldsymbol{c}} \sum_j\sum_y \Tilde{\mathscr{D}}\{d_j(y),g_j(y:\boldsymbol{c})\}\Big\}-\boldsymbol{c}^{k-1}
\end{equation}
where $k$ denotes the index of the current block, and $\Tilde{\mathscr{D}}$ denotes the surrogate function of the data fidelity term at $\boldsymbol{c}=\boldsymbol{c}^{k-1}$.

Then, the DECT image and DC output are stacked along the channel dimension as the input to the truncated U-Net. We also employ residual learning \cite{He2016} and group normalization \cite{Wu2018} to make training more efficient. The output of the $k$-th update block $\Gamma_k$ can be written as

\begin{equation}
    \Gamma_{\theta_k}(\boldsymbol{c}^{k-1})=\texttt{UNET}_k(\texttt{stack}(\mathcal{DC}(\boldsymbol{c}_{k-1},d), \boldsymbol{c}^{k-1}))+\boldsymbol{c}^{k-1}.
\end{equation}

Since the training process with intermediate loss functions has been observed to converge much faster than training with only the final loss, our network is trained under the supervision of the weighted sum of a set of intermediate loss functions as

\begin{equation}
    \arg\min_{\theta}\sum_{n=0}^{N-1} w_{e,n}\cdot ||\Gamma_{\theta_n}\circ\Gamma_{\theta_{n-1}}\dots\circ\Gamma_{\theta_{0}}(\boldsymbol{c}_{init})-\boldsymbol{c}_{truth}||,
\end{equation}
where $w_{e,n}\in[0,1]$ denotes the scalar that controls the backpropagation weight, $N$ denotes the number of stacked blocks, $e$ denotes the index of the current training epoch, ${\theta}=\{\theta_0,\theta_1…\}$ denotes the set of trainable parameters in all update blocks, $||\cdot||$  denotes the L2 norm, and $\circ$ denotes the function composition.

The GPU memory footprint is a critical design issue for 3D unrolled networks since the training process stores all feature maps for backpropagation, and 3D unrolled networks generate more and larger latent maps. This work mainly uses two techniques to reduce the GPU memory footprint. 

1) We partition the image volume into multiple small stacks consisting of 8 slices. The transmission data corresponding to each image stack is also truncated accordingly. Prior to passing to the DC layer, each stack is padded by 10 slices on each side to address the margin effect of helical CT reconstruction.

2) The double 3D convolutional layer with 3×3×3 kernels is substituted with a single 3D convolutional layer with 5×5×5 kernels. Compared to the original U-Net, the number of starting feature maps is also reduced from 64 to 40. 

These two modifications reduce the GPU memory footprint of the network by approximately 70\%.

\section{Results}
\subsection{Data Acquisition}

In this work, model-based reconstructed  images were taken as the reference. Five helical scans  at 90 and 140 kVp of a head-and-neck cancer patient were sequentially acquired by a Philips Brilliance Big Bore CT scanner at a 0.75 mm×16 collimation. These helical scans were then reconstructed by a joint statistical DECT algorithm, DEAM \cite{Zhang2018}, with the motion-compensation technique \cite{Ge2021} and a sufficient number of iterations. We selected four patient data as the training set and one as the test set. The dimension of the sinogram was 16×816×52800 with 40 rotations. The image size was 610×610×340, and the image resolution was 1×1×1.034 mm$^3$. The training set had 168 samples in total. Each SIR takes 45 hours on 4× NVIDIA V100 32 GB GPUs (400 iterations with 33 ordered subsets, plus 1000 iterations without ordered subset).
 
\subsection{Results}

We used a pretraining strategy to reduce the training time. Firstly, the first block of MB-DECTNet was trained individually for 2000 iterations. In this step, the reference update from the DC layer can be precomputed to save time. Then, the weights of the pretrained block were broadcasted to other blocks, and the entire network was trained end-to-end. The MB-DECTNet consisted of four updating blocks. 

\begin{figure}
  \centering
  \includegraphics[width=0.98\textwidth]{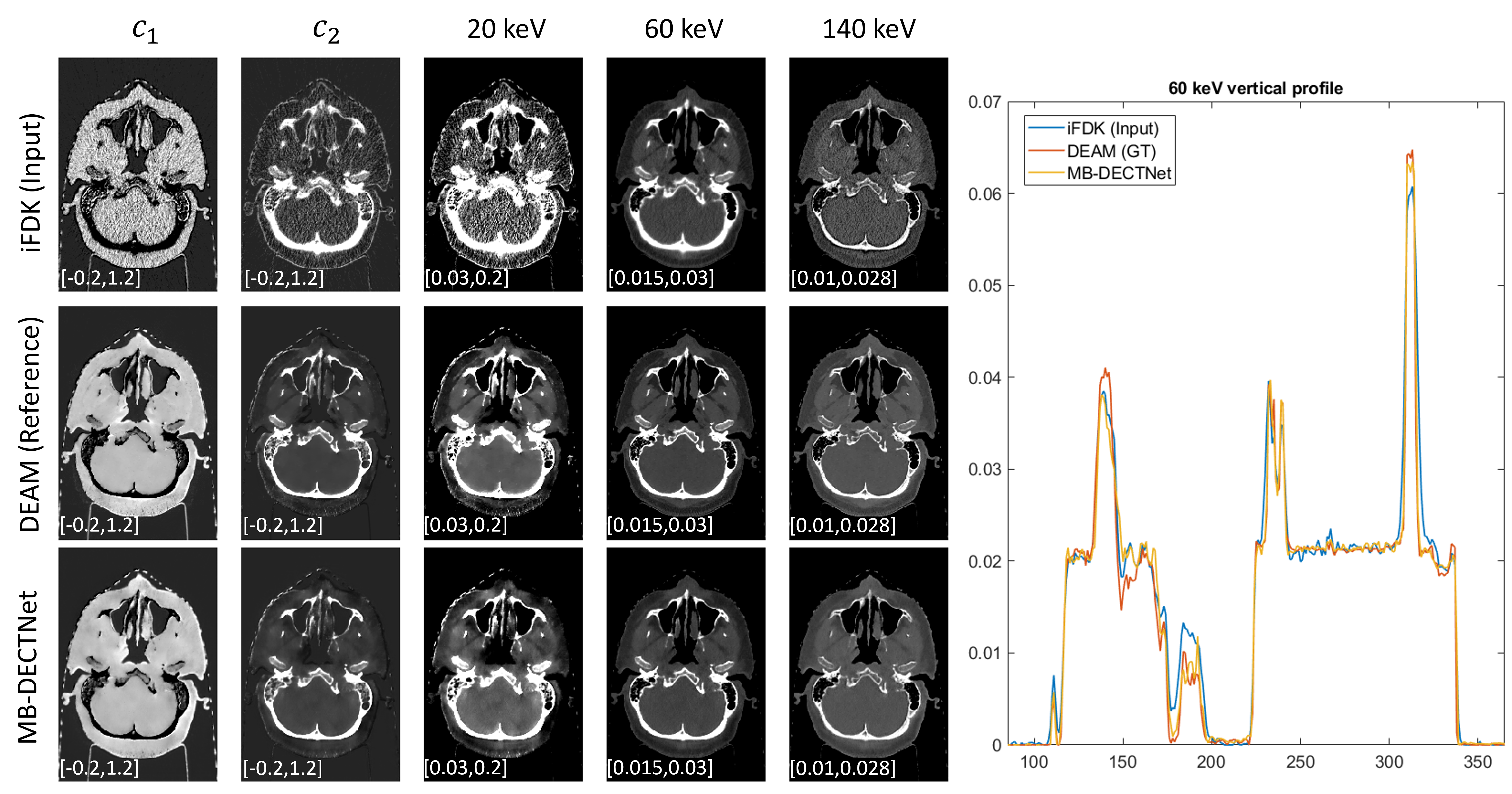}
  \caption{Images and middle vertical profiles for a selected slice in the test dataset.}
  \label{fig:result1}
\end{figure}

In the inference mode, MB-DECTNet generates a 610×610×340 image in approximately 350 seconds, which is 462-fold times shorter than the DEAM. Figure \ref{fig:result1} shows the image of a slice as an example of the inference result from MB-DECTNet, compared to iFDK (input) and DEAM (ground truth) images. iFDK refers to the 3D analytical DECT reconstruction that combines the FDK algorithm and the iterative filtered back-projection \cite{Yan2000}. MB-DECTNet image is much less noisy that the initial FDK estimate while preserving contrast and image sharpness. Indeed, the vertical profile suggests that MB-DECTNet improves the spatial resolution compared to the iFDK image.

\begin{figure}
  \centering
  \includegraphics[width=0.98\textwidth]{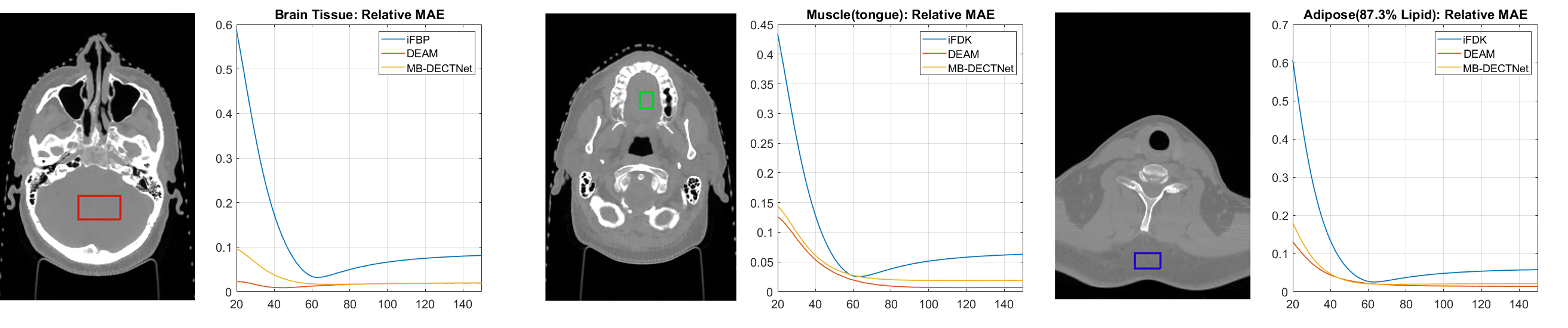}
  \caption{Plots of relative mean absolute error (RMAE) versus energy for selected regions of interest.}
  \label{fig:result2}
\end{figure}

More quantitatively, figure \ref{fig:result2} shows the relative mean absolute error (RMAE) for three different tissue ROIs relative to the LAC derived from their expected ICRU compositions \cite{White1987}, defined as

\begin{equation}
    \texttt{RMAE}(E)=\frac{\frac{1}{N}\sum_{x}|c_1(x)\mu_1(E)+c_2(x)\mu_2(E)-\mu(x,E)|}{\mu(x,E)}.
\end{equation}

MB-DECTNet shows its potential to estimate accurate LACs for energies at 20-150 keV, and the performance of MB-DECTNet is close to the performance of the selected DECT SIR, with a much lower computational cost.

\section{CONCLUSIONS}

To the best of our knowledge, it is the first time that a model-based unrolled network is proposed and trained end-to-end to estimate accurate basis components from dual-energy CT sinograms. Our proposed MB-DECTNet is capable of reducing noise and increasing resolution, and can be combined with numerous joint statistical DECT algorithms. The quantitative result shows that MB-DECTNet has the potential to estimate LAC accurately for energies from 20 to 150 keV as the selected traditional statistical algorithm with a much lower computational cost.

\acknowledgments 
 
This project is supported by R01 CA212638 from the United States National Institutes of Health. We thank the Siteman Cancer Center for their help in the acquisition of clinical data.

\bibliography{main} 
\bibliographystyle{spiebib} 

\end{document}